\documentclass[aps,prl,final,showpacs,twocolumn,nofootinbib,notitlepage]{revtex4-1}
\usepackage{graphicx,color}
\usepackage{dcolumn}
\usepackage{amssymb}
\usepackage{amsfonts}
\usepackage{color}
\usepackage[usenames,dvipsnames,svgnames,table]{xcolor}

\usepackage{epsfig}
\usepackage{bm}

\newcommand{\br}{{\bf r}}

\begin{document}

\title{Comment on "Tetraquarks as diquark-antidiquark bound systems"}

\author{M.~R.~Hadizadeh}
\affiliation{Institute of Nuclear and Particle Physics and Department of Physics
and Astronomy, Ohio University,
Athens, OH 45701, USA}

\date{\today}

\begin{abstract}
The author argues that the calculated masses of heavy tetraquarks obtained by solution of the spin-independent homogeneous Lippmann-Schwinger integral equation in a diquark-antidiquark picture reported by M. Monemzadeh et al., Phys. Lett. B {\bf741}, 124 (2015), are incorrect. 
We have reexamined all of the published results and we believe that not only the reported tetraquark masses for states with zero total angular momentum are incorrect, the reported masses for states with non-zero total angular momentum are quite misleading, because these states cannot be predicted by a spin-independent formalism.
\end{abstract}


\maketitle

In a recent letter by Monemzadeh et al. \cite{Monemzadeh-PLB}, the tetraquark bound state is studied as a two-body problem in a diquark-antidiquark picture. The nonrelativistic $s-$wave bound state of a diquark-antidiquark system with reduced mass $\mu_{D\bar D}$ and the binding energy $E_T$, interacting with pair force $V(r)$ can be described in configuration space by homogeneous Lippmann-Schwinger (LS) integral equation
\begin{equation} \label{LS}
\psi(r) = -\frac{2 \, \mu_{D\bar D}}{4\pi} \int \, d^3 \br' \, \frac{e^{-\kappa |\br - \br'|}}{|\br - \br'|} \, V(r') \, \psi(r'),
\end{equation}
where $\kappa= \sqrt{2\,\mu_{D\bar D}|E_T|}$. The corresponding equation in published letter \cite{Monemzadeh-PLB} (i.e. Eq. 2 and consequently Eqs. 4, 9 and 11) is missing a factor of $(2\pi)^{-1.5}$ which should be a typo, because our numerical analysis demonstrates that missing this factor leads to completely unreasonable tetraquark masses\footnote{To verify derivation of LS integral equation (\ref{LS}) and test our numerical algorithm and code, we have solved it for spin-independent Malfliet-Tjon Yukawa potential \cite{Malfliet-Tjon} and we have reproduced the same deuteron binding energy obtained from solution of LS equation in momentum space.}. 

 In order to study tetraquark bound states, the spin-independent part of the diquark-antidiquark potential given in Ref. \cite{Ebert-PLB634} is used, with the strong coupling constant $\alpha_s$ of Ref. \cite{ebert2011masses}
\begin{equation}
\label{alphas}
\alpha_s (\mu)= \frac{4\pi}{\beta_0} \frac{1}{\ln \left ({\frac{4\mu^2_{D\bar{D}} + M_0^2}{\Lambda^2}} \right)},
\end{equation}
with $\beta_0=11- \frac{2}{3} n_f$ ($n_f$ is the number of flavor quarks),
$M_0= 2.24 \sqrt{A} = 0.95\, \text{GeV}$ and $\Lambda=0.413 \, \text{GeV}$.

In Table \ref{Table.cqcq_cscs} we have listed our numerical results for $1s$ heavy charm tetraquarks for states with total angular momentum equal to zero.
As it is shown our calculated masses are all larger than the predicted results of Ref. \cite{Ebert-PLB634} which are obtained from solution of relativistic and spin-dependent LS integral equation in momentum space. 
Our results seem to be quite reasonable, because our numerical analysis indicates that the relativistic effects lead to a reduction in the calculated masses \cite{Hadizadeh-in-preparation}. 
Clearly the reported results in Ref. \cite{Monemzadeh-PLB} can not be trusted, because as it is shown in Table \ref{Table.cqcq_cscs}, not only $J=0$ states are not calculated correctly, they have reported the masses for non zero $J$ states, like $\frac{S\bar{A}\pm A\bar{S}}{\sqrt{2}}$ and $A\bar{A}$, which cannot be obtained in this formalism.
It is a serious challenge and the authors should clarify how this spin-independent formalism can distinguish different spin states and consequently the solution of spin-independent LS integral equation can predict the masses of tetraquarks with non zero total angular momentum.

\begin{table}[hbt]
\centering
\caption{Masses of charm diquark-antidiquark for $1s$ state with total angular momentum $J=0$ calculated from non-relativistic Lippmann-Schwinger equation. $S$ and $A$ denote the scalar and axial vector diquarks.}
\label{Table.cqcq_cscs}
\begin{tabular*}{\columnwidth}{@{\extracolsep{\fill}}ccccccc@{}}
\hline 
 Diquark content  & & Mass [GeV] &  \\  
 \hline
& & $cq\bar{c}\bar{q}$ &  \\  \cline{2-4}
 & present & Ref. \cite{Monemzadeh-PLB}  &  Ref. \cite{Ebert-PLB634} \\  \cline{2-4}
$S\bar{S}$ & 3.885  & 3.70314 & 3.812   \\
 $A\bar{A}$ & 4.013  & 3.83908 & 3.852  \\
\hline
& & $cs\bar{c}\bar{s}$ &  \\  \cline{2-4}
 & present & Ref. \cite{Monemzadeh-PLB}  &  Ref. \cite{Ebert-PLB634} \\  \cline{2-4}
   $S\bar{S}$ &  4.117  & 4.05390 &  4.051 \\
  $A\bar{A}$ &  4.250  & 4.09962 &  4.110 \\
\hline 
\end{tabular*}
\end{table}

Moreover, from numerical point of view the following issue should be considered.
Since a regularized form of the diquark-antidiquark potential is used to overcome the singularity of the confining potential at large distances, the calculated masses should be independent of the regularization cutoff $r_c$. In Ref. \cite{Monemzadeh-PLB} the authors have chosen the non-zero root of the potential as regularization cutoff and consequently the potential is fixed equal to zero for $r\ge r_c$. As we have shown in Table \ref{Table.reg-cut} clearly this cutoff is not high enough and in order to achieve the cutoff-independent results, converged with four significant digits, one needs to choose a cutoff at least equal to $5$ GeV$^{-1}$. Choosing a regularization cutoff equal to non zero root of the diquark-antidiquark potential leads to smaller tetraquark masses which are shown with bold numbers in Table \ref{Table.reg-cut}, even with the same regularization cutoff the reported masses in Ref. \cite{Monemzadeh-PLB} are different from our results.

We should mention that in our calculations we have used $100$ mesh points of Gauss-Legendre quadrature for integration over angle between $\br$ and $\br'$ with a linear mapping, whereas the integration over $r'$ is done by a hyperbolic-linear mapping with the sub-intervals $[0,\frac{r_c}{2}] \, + [\frac{r_c}{2},r_c] + [r_c,20]$ GeV$^{-1}$ using 150 mesh points to achieve the converged masses with four significant digits .


\begin{table}[hbt]
\centering
\caption{Masses of charm diquark-antidiquark for $1s$ state with total angular momentum $J=0$ as a function of regularization cutoff $r_c$. The bold numbers are calculated for values of $r_c$ which are non zero root of the diquark-antidiquark potential and numbers in parentheses are corresponding masses from Ref. \cite{Monemzadeh-PLB}.}
\label{Table.reg-cut}
\begin{tabular*}{\columnwidth}{@{\extracolsep{\fill}}ccccccc@{}}
\hline 
 $r_c$ [GeV$^{-1}$]  & & Mass [GeV] &  \\  
 \hline
& & $cq\bar{c}\bar{q}$ &  \\  \cline{2-4}
& $S\bar{S}$   & & $A\bar{A}$  \\
2.831 & 3.858 & & {\bf 3.987} (3.83908) \\
2.844 & {\bf 3.859} (3.70314) & & 3.987 \\
2.85 & 3.859 & & 3.988 \\
2.9 & 3.862 & & 3.991 \\
3.0 & 3.867  && 3.995 \\
3.5 & 3.879 & & 4.007 \\
4.0 & 3.883  & & 4.011 \\
5.0 & 3.885  & &  4.013 \\
6.0 & 3.885  & & 4.013 \\
\hline
& & $cs\bar{c}\bar{s}$ &  \\  \cline{2-4}
& $S\bar{S}$  & & $A\bar{A}$  \\
2.808 & 4.092 & & {\bf 4.226} (4.09962) \\
2.821 & {\bf 4.093} (4.05390) & & 4.227 \\
2.85 & 4.094 & & 4.228 \\
2.9 & 4.097 & & 4.231 \\
3.0 & 4.101 & &  4.235 \\
3.5 & 4.112 & & 4.245 \\
4.0 & 4.116 & &  4.248 \\
5.0 & 4.117 & &  4.250 \\
6.0 & 4.117  & & 4.250 \\
\hline
\end{tabular*}
\end{table}

Although the conclusions based on the incorrect results are misleading but we don't understand this part of their conclusion: "our results are in good agreement with the results derived from complicated relativistic methods and can be a good replacement for them". How a nonrelativistic spin-independent method can be an alternative to a relativistic spin-dependent method, even when the results of both methods be in good agreement?

In conclusion, our calculations demonstrate that all of the results published in recent letter by M. Monemzadeh et al. \cite{Monemzadeh-PLB}, for masses of tetraquarks with total angular momentum equal to zero, are incorrect and the letter has serious numerical problems. The correct results should be as results given in Table \ref{Table.cqcq_cscs}. 
 Moreover, beside these incorrect results, we have difficulty to understand how the authors have obtained the masses of tetraquarks for non zero total angular momentum in a spin-independent formalism!

\section*{Acknowledgments}
The author would like to thank V.O.~Galkin for providing the values of strong coupling constant for different tetraquark states. This work is supported by National Science Foundation under contract NSF-PHY-1005587
with Ohio University. Partial support was also provided by the Institute of Nuclear and Particle Physics at Ohio University.


\end{document}